\documentclass[aps,pra,showpacs,twocolumn,amsmath,amssymb,amsthm]{revtex4-1}
\usepackage{graphicx,bbm,enumerate}

\newcommand{\be}{\begin{equation}}
\newcommand{\ee}{\end{equation}}

\newcommand{\bra}[1]{{\langle #1 \vert}}

\newcommand{\ket}[1]{{\vert #1 \rangle}}

\newcommand{\ii}{ {\rm i} }

\newcommand{\RR}{\mathbb{R}}

\newcommand{\y}{{\rm y}}
\newcommand{\x}{{\rm x}}
\newcommand{\z}{{\rm z}}

\newcommand{\LL}{{\hat {\cal L}}}

\newcommand{\PP}{{\hat {\cal P}}}

\def\ad{{{\rm ad}\,}}
\def\one{\mathbbm{1}}

\def\PP{\hat{\cal P}}
\def\Pm{\mathbb{P}}

\def\Tm{\mathbb{T}}
\def\LL{\hat{\cal L}}
\def\DD{\hat{\cal D}}

\begin{document}

\title{Generic examples of $\Pm \Tm$-symmetric qubit (spin $1/2$) Liouvillians}
\author{Toma\v{z} Prosen}
\affiliation{Department of Physics, FMF,  University of Ljubljana, Jadranska 19, 1000 Ljubljana, Slovenia}
\date{\today}

\begin{abstract}
We outline two general classes of examples of $\Pm\Tm$-symmetric quantum Liouvillian dynamics of open many qubit systems, namely interacting hard-core bosons (or more general $XYZ$-type spin 1/2 systems) with, either (i) pure dephasing noise,  or (ii) having solely single particle/spin injection/absorption incoherent processes.
\end{abstract}

 
\maketitle
The concept of PT symmetry has been introduced \cite{bender} as a mathematical framework for studying non-Hermitian operators with symmetric (e.g. real) spectra. In recent years, PT symmetry has been studied extensively, both theoretically \cite{znojil,ali,fleischmann,west,schomerus} and experimentally in the context of optics \cite{christo,christo2} and electric circuits \cite{lrc}. Very recently, a formally analogous concept has been proposed \cite{prl} on the level of master symmetries of quantum master equations \cite{lindblad,breuer,zoller} (Liouville equations) describing open quantum systems. It has been shown that the existence of
quantum $\Pm\Tm$ symmetry generically implies the existence of a spontaneous symmetry breaking transition, such that for sufficiently weak coupling to the environment all coherences (off-diagonal matrix elements of the system's density matrix in the energy eigenbasis) decay with the same (uniform) damping rate.

 The aim of this Brief Report is to provide two general classes of practically interesting examples to a recent construction of quantum Liouvillian $\Pm\Tm$ symmetry \cite{prl}. We use notation exactly as introduced in Ref.~\cite{prl} and refer to definitions stated there.

We start by proving a simple observation which seems to be crucial for a construction of general examples of $\Pm\Tm$-symmetric Liouvillian systems:
\bigskip

\noindent
{\bf Lemma.}
The Liouvillian flow is $\Pm\Tm$-symmetric, if the following conditions are fulfilled:
\begin{enumerate}[(i)]
\item The parity superoperator can be represented as
\begin{equation}
\PP\rho = U \rho W,
\nonumber
\end{equation} 
where $U,W\in {\cal B}({\cal H})$ are two unitary operators satisfying $$U^2=W^2=\one,$$ and $$[H,U]=[H,W]=0.$$
\item $\exists$ $M \times M$ real orthogonal reflection matrix $Z \in {\rm O}(M,\RR)$, satisfying $Z^2 = \one_{M}$, such that
\begin{eqnarray*} 
U L_m &=& -\sum_{m'=1}^M Z_{m,m'} L_{m'}^\dagger U, \\
W L_m &=& \sum_{m'=1}^M  Z_{m,m'} L_{m'}^\dagger W. 
\end{eqnarray*} 
\item $\{L_m,L_m^\dagger\} = c_m \one$, for some $c_m \in \RR$.
\end{enumerate}

\noindent
{\bf Proof.}
Using (iii), $\DD'$ can be written as 
\begin{equation}
\DD'\rho = \sum_m \left(2 L_m \rho L_m^\dagger - \frac{1}{2}\{ [L^\dagger_m,L_m],\rho\}\right),
\end{equation} 
while in the Hilbert-Schmidt metric
\begin{equation}
(\DD')^\dagger \rho = \sum_m \left(2 L^\dagger_m \rho L_m - \frac{1}{2}\{ [L^\dagger_m,L_m],\rho\}\right).
\end{equation} 
Then, using (ii), 
\begin{eqnarray}
\DD' \PP \rho &=& \DD'(U\rho W) \nonumber \\ &=&- U \sum_m \left(2 L_m^\dagger \rho L_m - \frac{1}{2}\{[L^\dagger_m,L_m],\rho\}\right)W \nonumber	 \\ &=& -\PP (\DD')^\dagger \rho.
\end{eqnarray} 
Finally, using (i), we have also 
\begin{equation}
(\ii\,\ad H)\PP = \PP(\ii\,\ad H) = -\PP (\ii\,\ad H)^\dagger,
\end{equation} 
and therefore 
\begin{equation}
\LL'\PP = -\PP(\LL')^\dagger,
\end{equation} 
i.e., definition of $\Pm\Tm$ symmetry (Eq. (8) of \cite{prl}). $\Box$

\bigskip

One, quite restricted example of application of $\Pm\Tm$-symmetric quantum Liouvllian dynamics has been provided in Ref. \cite{prl}. Here we show that applications are in fact quite abundant and should not be difficult to realize in experimentally accessible situations.

\bigskip
\noindent {\bf Example 1}.
Consider $n$ of spins $1/2$, or qubits, described by Pauli matrices $\sigma^{\rm x,y,z,\pm}_j$, $j=1\ldots n$.
Take an arbitrary {\em two-spin} Hamiltonian with a {\em longitudinal} external field
\begin{equation}
H = \sum_{j < k} (J^\x_{j,k} \sigma^\x_j \sigma^\x_k + J^\y_{j,k} \sigma^\y_j \sigma^\y_k + J^\z_{j,k} \sigma^\z_j \sigma^\z_k) + \sum_j h_j \sigma^\x_j,
\end{equation}
and up to $M=n$ local Hermitian `dephasing' Lindblad operators
\begin{equation}
L_j = \gamma_j \sigma^\z_j,\quad j=1\ldots n, 
\label{eq:lind1}
\end{equation}
The interaction strengths $J^{\x,\y,\z}_{j,k}$, magnetic field strengths $h_j$, and local dephasing rates $\gamma_j$ can be completely arbitrary.
Liouvillian dynamics defined with respect to the Hamiltonian $H$ and Lindblad operators $\{ L_j\}$ is $\Pm\Tm$ symmetric, according to the Lemma above, if we take
\begin{equation}
U = \prod_{j=1}^{n} \sigma^\x_j ,\qquad W = \one.
\end{equation}
In case $J^\x_{j,k} \equiv J^\y_{j,k}$, $h_j\equiv 0$, the model represents a general interacting system of hard-core bosons with zero (or constant) chemical potential.
Then, the Lindblad channels (\ref{eq:lind1}) indeed model the pure dephasing noise.

\bigskip
\noindent {\bf Example 2}.
Consider now a slightly more restriced Hamiltonian, again for  $n$ of spins $1/2$, 
\begin{equation}
H = \sum_{j < k} (J^\x_{j,k} \sigma^\x_j \sigma^\x_k + J^\y_{j,k} \sigma^\y_j \sigma^\y_k + J^\z_{j,k} \sigma^\z_j \sigma^\z_k),
\end{equation}
and up to $M=2n$  Lindblad operators which represent incoherent local spin/particle absorption/injection
\begin{equation}
L_{2j-1} = a_j \sigma^{\rm +}_j,\qquad L_{2j} = b_j \sigma^{\rm -}_j,\quad j=1\ldots n. 
\end{equation}
Again, the interaction matrices $J^{\x,\y,\z}_{j,k}$ and the spin-flipping rates  $a_j,b_j$ can be arbitrary.
The Lemma can now be implemented with
\begin{equation}
U = \prod_{j=1}^{n} \sigma^\y_j,\qquad W = \prod_{j=1}^{n} \sigma^\x_j.
\end{equation}
In the $XXZ$-like case $J^\x_{j,k}\equiv J^\y_{j,k}$ the model represents an open interacting hard-core boson model. For nearest-neighbor interaction corresponding to one dimension (i.e., a chain), and if incoherent processes are only at the ends,  $a_j=b_j=0$, $2\le j\le n-1$, the model also describes an open version of the so-called $t-V$ model of spinless fermions.

\bigskip

In both examples above, the conditions (i-iii) of the Lemma are straightforward to check with the reflection matrix being trivial $Z=\one_M$. Besides the necessary conditions $[H,U]=0$, $[H,W]=0$, implied by (i), we may or may {\em not} have $[U,V]=0$, hence the number of spins $n$ in our examples need not be even.

Provided the parameters in the Hamiltonian can be chosen such that both, the energy spectrum and the energy-difference (frequency) spectrum are non-degenerate, 
one should observe the spontaneous $\Pm\Tm$ symmetry breaking transition of Liouvillian decay rates while increasing the noise/dissipation strength \cite{prl}, and below the transition point all the coherences should decay with a uniform rate. It is reasonable to expect that both classes of examples should be relevant for experiments with ultra-cold atom systems.

We close this note by pointing out the fact that our explicit construction (Lemma above) mend the statement on the symmetry of the dissipator-perturbation matrix $V$ made in Eq.~(14) of Ref. \cite{prl}, which in general seems to be in-conclusive. The symmetry of $V$ is in fact needed in order to establish a non-trivial sponaneous $\Pm\Tm$ symmetry breaking transition (i.e., nonvanishing of $\gamma_{\rm PT}$).
In order to prove $V_{j,k}=V_{k,j}$, we only need to show $\sum_m |\bra{\psi_j}  L_m \ket{\psi_k}|^2 = \sum_m |\bra{\psi_k}  L_m \ket{\psi_j}|^2$. Indeed, due to the condition (i) of the Lemma
 the eigenvectors $\ket{\psi_j}$ of $H$ can be chosen to be simultaneously eigenvectors of $W$, say $W \ket{\psi_j} = \omega_j \ket{\psi_j}$, with $\omega_j \in\{\pm 1\}$, 
so we have 
\begin{eqnarray*}
 &&\sum_m |\bra{\psi_j}  L_m \ket{\psi_k}|^2 \\
 &=& \sum_m |\bra{\psi_j} W L_m \ket{\psi_k}|^2 \\
 &=&  \sum_{m,m',m''} Z_{m,m'} Z_{m,m''} \bra{\psi_j} L_{m'}^\dagger W \ket{\psi_k}\overline{\bra{\psi_j} L_{m''}^\dagger W \ket{\psi_k} } \\
 &=& \sum_{m'} |\bra{\psi_j}L^\dagger_{m'}\ket{\psi_k}|^2 = \sum_{m} |\bra{\psi_k}L_{m}\ket{\psi_j}|^2,
 \end{eqnarray*}
 since $Z$ is orthogonal, $Z^T Z = \one_M$.  $\Box$


\begin{thebibliography}{10}
\bibitem{bender} C.~M.~Bender and S.~Boettcher, Phys. Rev. Lett. {\bf 80}, 5243 (1998);
C.~M.~Bender, Rep. Prog. Phys. {\bf 70}, 947 (2007).

\bibitem{znojil} M.~Znojil, Phys. Lett. A {\bf 259}, 220 (1999); G.~Leval and M. Znojil, J. Phys. A: Math. Gen. {\bf 33}, 7165 (2000).

\bibitem{ali} A.~Mostafazadeh, J. Math. Phys. {\bf 43}, 3944 (2002); Phys. Rev. Lett. {\bf 102}, 220402 (2009); Int. J. Geom. Meth. Mod. Phys. {\bf 7}, 1191 (2010).

\bibitem{fleischmann} O.~Bendix, R.~Fleischmann, T.~Kottos, and B.~Shapiro, Phys. Rev. Lett. {\bf 103}, 030402 (2009).

\bibitem{west} C.~T.~West, T.~Kottos, and T. Prosen, Phys. Rev. Lett. {\bf 104}, 054102 (2010).

\bibitem{schomerus} H.~Schomerus, Phys.~Rev.~Lett.~{\bf 104}, 233601 (2010); Phys. Rev. A {\bf 83}, 030101(R) (2011).

\bibitem{christo} K.~G.~Makris, R.~El-Ganainy, D.~N.~Christodoulides and Z.~H.~Musslimani, Phys. Rev. Lett. {\bf 100}, 103904 (2008);
A.~Guo, {\em et al.}, Phys. Rev. Lett. {\bf 103}, 093902 (2009).

\bibitem{christo2} Z.~Lin, H.~Ramezani, T.~Eichelkraut, T.~Kottos, H.~Cao, and D.~N.~Christodoulides, Phys. Rev. Lett. {\bf 106}, 213901 (2011).

\bibitem{lrc} J.~Schindler, A.~Li, M.~C.~Zheng, F.~M.~Ellis, and T.~Kottos, Phys. Rev. A {\bf 84}, 040101 (2011);
Z.~Lin, J.~Schindler, F.~M.~Ellis, and T.~Kottos, {\em ibid} {\bf 85}, 050101 (2012);
H.~Ramezani, J.~Schindler, F.~M.~Ellis, U.~GŸnther, and T.~Kottos, {\em ibid} {\bf 85}, 062122 (2012).

\bibitem{prl} T. Prosen, Phys. Rev. Lett. {\bf 109}, 090404 (2012).

\bibitem{lindblad} G.~Lindblad, Commun. Math. Phys. {\bf 48}, 119 (1976);
V.~Gorini, A.~Kossakowski, and E.~C.~G.~Sudarshan, J. Math. Phys. {\bf 17}, 821 (1976).

\bibitem{breuer} H.-P. Breuer and F. Petruccione, {\em The theory of open quantum systems}, (Oxford University Press,  New York 2002).

\bibitem{zoller} C.~W.~Gardiner and P.~Zoller, {\em Quantum Noise: A Handbook of Markovian and Non-Markovian Quantum Stochastic Methods with Applications to Quantum Optics},
(Springer-Verlag, Berlin Heidelberg 2004).

\end{thebibliography}
\end{document}